# A METHODOLOGY FOR CALCULATING THE LATENCY OF GPS-PROBE DATA

**Zhongxiang Wang**
Graduate Student
Department of Civil & Environmental Engineering
University of Maryland
Email: zxwang25@umd.edu

**Masoud Hamedi**[*]
Senior Research Scientist
Center for Advanced Transportation Technology
University of Maryland
5000 College Ave
College Park, MD 20740
Email: masoud@umd.edu | Phone:301-405-2350

**Stanley Young**
Advanced Transportation and Urban Scientist
National Renewable Energy Laboratory
Email: stanley.young@nrel.gov

Word count:  4,628 words text + (3 tables +8 figures) x 250 words (each) = 7,378 words

Submission Date: Aug 1[st], 2016

## ABSTRACT

Crowdsourced GPS probe data has been gaining popularity in recent years as a source for real-time traffic information. Efforts have been made to evaluate the quality of such data from different perspectives. A quality indicator of any traffic data source is latency that describes the punctuality of data, which is critical for real-time operations, emergency response, and traveler information systems. This paper offers a methodology for measuring the probe data latency, with respect to a selected reference source. Although Bluetooth re-identification data is used as the reference source, the methodology can be applied to any other ground-truth data source of choice (i.e. Automatic License Plate Readers, Electronic Toll Tag). The core of the methodology is a maximum pattern matching algorithm that works with three different fitness objectives. To test the methodology, sample field reference data were collected on multiple freeways segments for a two-week period using portable Bluetooth sensors as ground-truth. Equivalent GPS probe data was obtained from a private vendor, and its latency was evaluated. Latency at different times of the day, the impact of road segmentation scheme on latency, and sensitivity of the latency to both speed slowdown, and recovery from slowdown episodes are also discussed.

# INTRODUCTION

Accurate and timely data is a vital component of any Intelligent Transportation System. In recent years, proliferation of location-aware internet connected devices has enabled private sector to use crowd sourcing technics for providing network wide real-time travel time and speed data for traffic management applications. This has resulted in traffic data services that report speed and travel time in real-time. This data in turn is used by private industry for traveler information and routing, and increasingly by public entities as a replacement for field data collection and to expand observability of roadway conditions network wide. The I-95 Corridor Coalition's Vehicle Probe Project has successfully integrated third party data of this nature, commonly referred to as probe data, for a number of public agency applications. Initial concerns about accuracy were addressed by a comprehensive validation program that compared probe industry reported speeds and travel times with those from a sensor-based reference source. Real-time applications are also sensitive to the latency, that is the time delay between actual field conditions, such as a major slowdown, and when it is reflected in the traffic data stream. Appropriate method to benchmark latency is currently lacking, and is the focus of this paper.

Consumer electronics are finding an ever-increasing role in our everyday lives. A majority of these devices are also equipped with a point-to-point networking protocol commonly referred to as Bluetooth. Bluetooth enabled devices can communicate with other Bluetooth enabled devices anywhere from one meter to about 100 meters. This variability in the communications capability depends on the power rating of the Bluetooth sub-systems in the devices. The Bluetooth protocol uses an electronic identifier, or tag, in each device called a Machine Access Control address, or MAC address for short. The MAC address serves as an electronic nickname so that electronic devices can keep track of who's who during data communications. In principle, the Bluetooth traffic monitoring system calculates travel times by matching public Bluetooth MAC addresses at successive detection stations. Bluetooth data has been accepted by the industry as an accurate and economic solution for collecting ground-truth travel time data. More details on using Bluetooth sensors for freeway travel time data collection is discussed in (1).

Quality and accuracy of the GPS probe data has been validated mostly compared to the Bluetooth data by many researchers. Average Absolute Speed Error and Speed Error Bias are among the quality measures. However, not much effort has been made to quantify latency of the probe data as an indicator of its punctuality.

In the context of travel time data, latency can be defined as the difference between the time the traffic flow is perturbed and the time that the change in speed is reflected in the data. When using Bluetooth data as ground-truth, latency is measured by observing the time difference between the onset of a slowdown as reported by Bluetooth traffic monitoring, and the time that it is reported by the GPS probe data. A graphical representation is shown in FIGURE 1. The time shift between probe data and Bluetooth data, which is marked with orange arrow, is the latency of GPS probe data.

Latency associated with the GPS probe data originates from several sources, and is unavoidable to some extent. Figure 2 shows a conceptual framework for generating GPS probe data. Every second, millions of GPS tracks are being collected and there is a delay from the time that an observation is made on the field, to the time that it is transmitted through a communication medium to the data collection server. The GPS data is blended with other data sources such as historical, incident and weather data and goes through data fusion and filtering engines, which takes some time. The fused data is then packaged in predetermined formats and is injected into live feeds for consumer consumption. The applications running on the user side pull the data. So by the time that data is available to the real-time applications, traffic dynamics might have changed on the field. It is important to have a good understanding of the delay, in order to tune the real-time applications. The term 'latency' is used in industry literature in many contexts related to various steps along the processing chain as shown in FIGURE 2. For the purpose of this research, latency is defined as "the difference between the time the traffic flow is perturbed and the time that the change in speed is reflected in the data". This represents a system latency – not any specific step in the chain. This is the latency that the proposed methodology addresses.

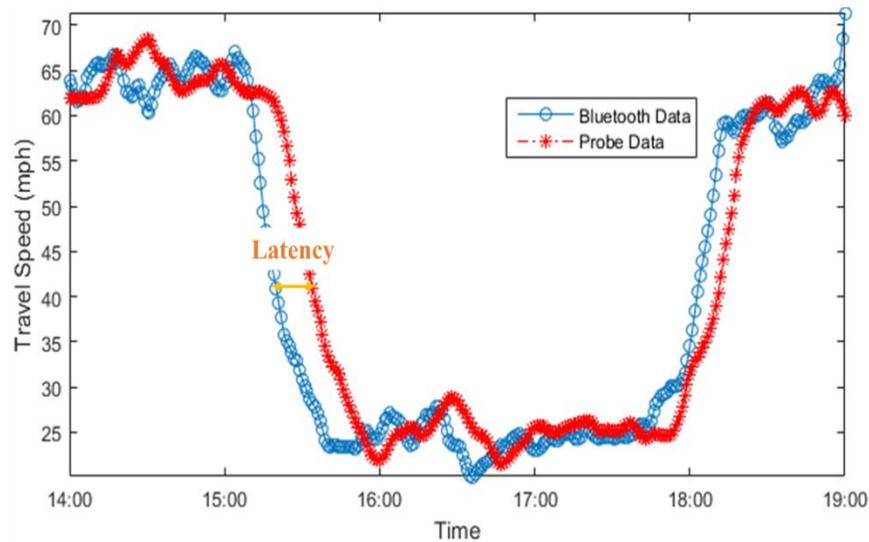

**FIGURE 1 Measurement of latency**

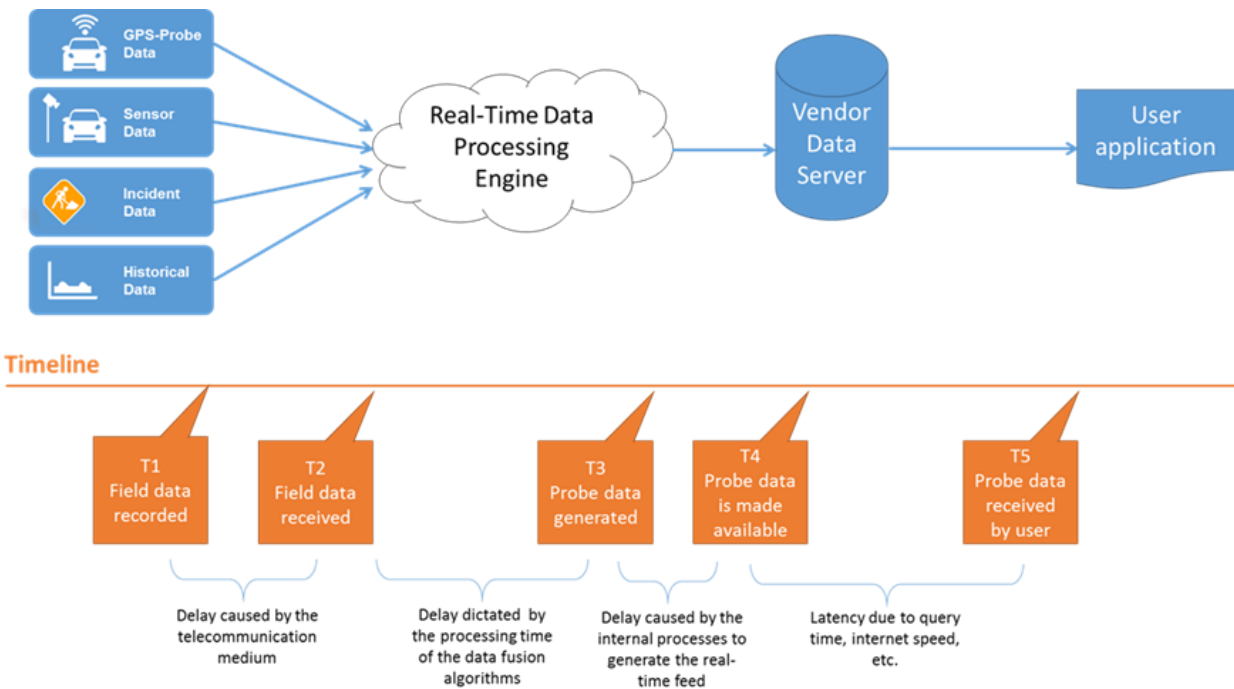

**FIGURE 2 GPS-probe data processing flow chart**

The paper is organized as following: First a brief review about research on probe data latency is present. Then a methodology, including data processing steps and an iterative matching procedure with three fitness objectives for calculating latency is presented. A case study based on extensive field collected Bluetooth and GPS probe data is conducted to test the proposed methodology. In depth analysis of the case study results including sensitivity of latency to spatial and temporal parameters are presented. Finally, main takeaways and direction for future research are summarized.

**LITERATURE REVIEW**
Latency measurement for real-time travel time data is a relatively untouched research topic. Haghani et al. (2) estimated that lag time for GPS-probe data is less than or equal to eight minutes, however they did not specify the source for that statement. Liu et al. (3) observed a clear latency for reporting GPS-based data as well. Chase et al. (4) further stated that GPS-probe data has greater latency when travel speed recovers after peak period, than the beginning of a peak period. None of these papers proposed a detailed quantitative latency measurement methodology.

Kim and Coifman (5) measured the latency for GPS-probe data compared to loop detector. They calculated the correlation coefficient, which significantly depends on the covariance of original time-series speed data and shifted time-series speed data. The results show that the average latency for GPS-probe data is 6.8 minutes, and it could exceed 10 minutes in many situations. However, loop detector can only report spot travel speed whereas GPS probe data is reported on standardized segments, typically Traffic Message Control (TMC) industry standard segments. The speed reported is more related to the space mean speed across a TMC segment rather than a spot speed from loop detectors. Moreover, Kim and Coifman shifted the GPS-probe data in 10 second increments. However, the granularity of commercially available GPS probe data is one minute or more.

**METHODOLOGY**
The methodology used in this paper for quantifying latency of the probe data with respect to Bluetooth data involves multiple steps. The following sections provide a brief description for each step. It would be good to have a succinct overview of the steps here such as – data preparation, filtering, data interpolation and smoothing. The method used compares a reference data source (Bluetooth re-identification data) that directly samples travel time on a segment basis, converts the travel times to speed measures, and then compares the speed to that reported by probe data sources. An error metric is calculated between the Bluetooth reference data and the industry probe data. The probe data is then time shifted until minimum error is achieved. The time-shift that creates the minimum error is the latency of the data. Various steps address data preparation (outliers, smoothing, etc.) but the basic approach is as described.

**Bluetooth Data Preparation**
Bluetooth sensors store the MAC ID of the detected Bluetooth devices along with their detection time in a removable memory card. The collected data are downloaded to a server for processing at the end of deployment. The MAC addresses for all devices that are detected between two consecutive sensors are matched to develop a sample of travel time for that particular segment of the roadway. The reader is reminded that travel time and speed are inversely related and throughout this paper, they have been used interchangeably. It should be noted that the conversion to speed is based on the measured distance between sensor locations. In order to establish the ground truth for

Wang, Hamedi, Young                                                                                                       6travel time, individual observations must be aggregated in specified time intervals which in this paper are assumed to be equal to one minute. It must also be noted that the detection time of the second sensor is used as the time label for the individual observations. Space mean speed for each interval is equal to summation of travel time of all observations divided by product of number of observations by the segment length. It should be noted that the time reference used in this data is the time when vehicles are re-identified at the downstream sensor (as opposed to the upstream sensor, or the mean time of the upstream and downstream.) This will be discussed in more detail later.

**Data Filtering**
Due to the nature of traffic movement, some data points obtained in the matching stage are, in fact, unacceptable due to several reasons. For example, if after detection by the first sensor, a driver pulls over to replace a flat tire, after reaching the second sensor a travel time observation will be generated that is not a valid representation of average traffic pattern. In summary, data samples with the following characteristics must be identified:

- Observations with unreasonably low speeds,
- Observations in a particular time interval that are far from the average of the rest of the speeds observed in the same time interval to avoid erratic variations, and,
- Presence of a small number of observations in a time interval that is not enough to establish a reliable "ground truth" speed.

To address each of the mentioned potential problems, a series of filters were sequentially applied to the pool of unfiltered observations that result from the matching step. Variations in speed observations are considered to identify outlier speed observations. To that end, all observations corresponding to each of the time intervals for which we have Bluetooth observations are identified and the average and standard deviation of the speeds in those time intervals are calculated. Observations that correspond to speeds falling within ±1.5 times the standard deviation are kept and the rest are discarded. Assuming a normal distribution for the observations around the mean, this approach translates into keeping nearly 87 percent of the data. To ensure that the variability among speed observations inside a given time interval is within a reasonable level, the coefficient of variations (COV) of Bluetooth speed observations in each time interval that survive the previous step is estimated, then time intervals that have a COV greater than 1 are excluded and their corresponding observations are discarded from further consideration in the ground truth estimation process. More details on Bluetooth data matching and filtering are reported in (1). Many of the assumptions and procedures for filtering are based on a speed distribution with little short-term volatility. If two distinct speed distributions occur on a roadway (as is sometimes experienced, particularly on signalized roadways, much-less so on freeways), the methodology should be used with caution.

**Interpolation**
The third step is to interpolate so that the time-series data would have continuous coverage with one aggregated data point per each minute. The average of neighboring observations is considered to be the travel speed for the missing interval:

$$s_{t+i} = s_t + \frac{i}{n+1}(s_{t+n+1} - s_t) \tag{1}$$

Assuming that there are *n* consecutive missing data points starting from time *t*, equation (1) is used to fill the gap between time *t* and *t+n*. In order to preserve the consistency and integrity of the original data, it was chosen to use the gap filling formula only when not more than five consecutive data points are missing. Any sample data with larger than five-minute data gaps were excluded from the analysis. The interpolation procedure was applied to both Bluetooth and GPS probe data sets.

**Smoothing**
The final step is smoothening of the raw data. This procedure minimizes sudden sharp spikes in the general data trend caused by randomness of traffic speeds. Smoother curves allow comparison of the dominant pattern of the data curves, in order to calculate horizontal offset corresponding to the time gap. The filter function *filtfilt* in Matlab used to conduct the smoothing is based on rational transfer function, proposed by Oppenheim, Ronald and John (6), which is shown as in Equation 2.

$$Y(z) = \frac{b(1) + b(2)z^{-1} + \ldots + b(n_b+1)z^{-n_b}}{1 + a(2)z^{-1} + \ldots + a(n_a+1)z^{-n_a}} X(z) \tag{2}$$

where $n_a$ is the feedback filter order and $n_b$ is the feedforward filter order. When $a(2)$ to $a(n_a+1)$ are all zeros, this function degenerates into weighted moving average. $b(1)$, …, $b(n_b+1)$ are the weights for each data in the moving window. The moving average function becomes as the following function.

$$y(k) = w(1)x(k) + w(2)x(k-1) + \ldots + w(n)x(k-n+1) \tag{3}$$

$y(k)$ is the smoothed data at time $k$, $w(1)$ is the weight for its corresponding data. The moving time window is set to be five minutes. Each smoothed data point will be the summation of its weighted original data and weighted previous five minutes data points. One concern with smoothing is that it will shift the curve backward and therefore introduce an additional delay comparing to original data and as such the measured latency would be artificially higher than what is should be. To solve this problem, smoothing is applied to the raw data twice. First forward smoothing is done by moving the smoothing window forward, and then the smoothing is applied backwards with the same weight parameters which compensates the artificial latency.

**Latency Measurement**
Calculating the latency is an iterative process with a time offset, starting from a lower bound and repeating until reaching the preset upper bound. The underlying assumption is that probe data has the latency as explained in the introduction. The methodology to measure the latency is to find the best time shift which results in maximum match of Bluetooth curve and probe data curve. Three different fitness objectives are applied: Absolute Vertical Distance (*f1, AVD*), Square Vertical Distance (*f2, SVD*), and Correlation (*f3, COR*). Absolute Vertical Distance (*f1*) is the absolute value of the subtraction of Bluetooth travel speed and probe data travel speed over the desired measuring period. Square Vertical Distance (*f2*) is the square of that subtraction, which gives more weights to the points that have bigger difference. Correlation (*f3*) is a statistical representation of the linear relationship between two curves (7), which is defined as follows:

$$corr(S_t^{BT}, S_{t-latency}^{probe}) = \frac{E[(S_t^{BT} - \mu_{S_t^{BT}})(S_{t-latency}^{probe} - \mu_{S_{t-latency}^{probe}})]}{\sigma_{S_t^{BT}} \sigma_{S_{t-latency}^{probe}}} \quad (4)$$

Where $S_t^{BT}$ is the Bluetooth travel speed without shifting, $S_{t-latency}^{probe}$ is probe data travel speed shifted by the latency, E is the expected values, $\mu_{S_t^{BT}}$, $\mu_{S_{t+latency}^{probe}}$ are expected values of $S_t^{BT}$ and $S_{t-latency}^{probe}$ and $\sigma_{S_t^{BT}}$, $\sigma_{S_{t-latency}^{probe}}$ are the standard deviations of two curves.

The methodology starts from the lower bound (no latency, which means time offset is 0), increases time offset by 1 minute for each iteration, and then calculates all three fitness objectives. After reaching the upper bound for the time offset, the offset that results in best fit over all iterations is considered to be the latency. The formulation to find the shift distance which provides the most overlapping data is:

$$\min f1 = \sum_{t=1}^{n} \left| S_t^{BT} - S_{t-latency}^{probe\ data} \right|$$

$$\min f2 = \sum_{t=1}^{n} \left( S_t^{BT} - S_{t-latency}^{probe\ data} \right)^2 \quad (5)$$

$$\min f3 = corr(S_t^{BT}, S_{t-latency}^{probe\ data})$$

$lb \leq latency \leq ub$

To measure the latency of the probe data, both travel speed and travel time can be used. In order to make the objective metric to the segment length, this paper uses the former. If the Bluetooth data and GPS-probe data curves show the exact same pattern, shifting the probe curve will eventually result in a perfect match with zero vertical distance and correlation equals to 1. However, it is very unlikely due to random nature of traffic movement and also instrumentation error in both Bluetooth and probe technologies. The speed observations show less fluctuation during peak periods and heavy congestion conditions.

**CASE STUDY**
In order to test this methodology, two freeway corridors in South Carolina are selected. The first section is a 7.07-miles-long segment on I-85, from Exit 48 (US-276) to Exit 54 (Pelham Rd), which is shown in FIGURE 3 The second section is a 4.67-miles-long segment on I-26 from Exit 103 (Harbison Blvd) to Exit 108 (Bush River Rd), which is shown in FIGURE 3 The data was collected for both westbound and eastbound for I-26 and both northbound and southbound for I-85 from December 3, 2015 to December 15, 2015. These two paths are freeway where traffic would not be interrupted by traffic signals. Small blue dots show the location of Bluetooth sensors. Traffic Message Channel (TMC) codes used by probe data vendors to report data are also shown on the map. To improve quality of Bluetooth data, two sensors are deployed at each point, with one sensor on the outermost shoulder of each direction. In total 18 Portable Bluetooth sensors were deployed. Bluetooth sensors are marked by red capital letters, and a directional Bluetooth segment consists of data obtained from beginning and end sensor (i.e segment AB is the road segment starting from A and ending at B). GPS probe data was used in this study was acquired from a private vendor, in one minute granularity.

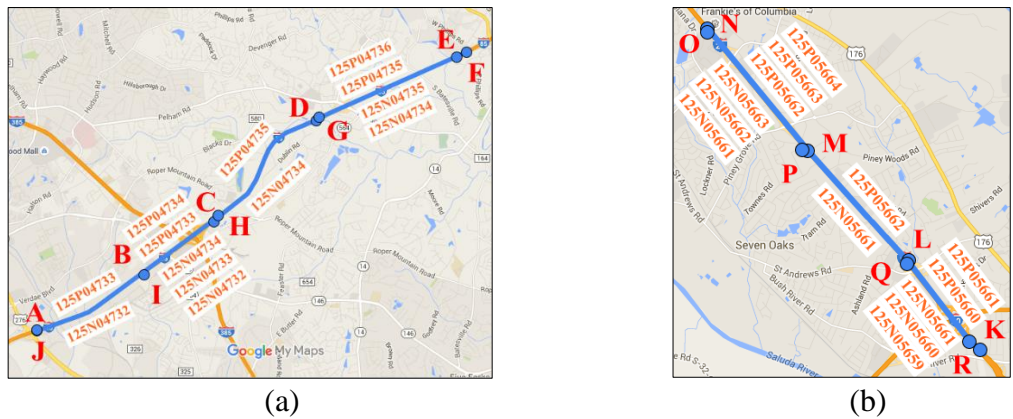

(a)                                                                        (b)

**FIGURE 3 Selected study area**

Then data is smoothed by the weighted moving average method discussed earlier. The moving window is set to be five minutes. Only previous data points are considered. Equation 6 depicts the weights and parameters used for smoothing. An example of smoothing of Bluetooth data is graphically shown in FIGURE 4(a) and similar for smoothing of GPS-probe data at FIGURE 4(b). The weights arithmetically decrease with respect to the increase of the time difference from the smoothing data point to the previous data point.

$$y(k) = 0.33x(k) + 0.27x(k-1) + 0.20x(k-2) + 0.13x(k-3) + 0.07x(k-4) \qquad (6)$$

After applying the latency measurement methodology, the latency is calculated and visualized on comparative graphs. FIGURE 4(c) is the original travel speed comparison between Bluetooth data and GSP-probe data and FIGURE 4 (d) depicts the same comparison after compensating latency (5 mins in this scenario) for GPS-probe data. It must be iterated that this number is only for one slowdown episode, on one segment of the road. In order to have a good understanding of the latency, it is important to apply the methodology to a large number of cases on different road segments, which is conducted later in the study

**Latency at Peak Periods**
To test the methodology on all segments, 32 morning peak period showdown episodes and 45 afternoon peak period showdown episodes are observed and identified. All other observations that happened on weekends and off-peak, or with data gap, or with different pattern between Bluetooth data and probe data, or the ones that did not exhibit clear slowdown pattern were excluded. The corresponding vendor TMC segments were selected and assigned to the location of Bluetooth sensors. If the Bluetooth sensor location does not match the exact TMC endpoint, the TMC is assigned to two adjacent segments based on the length of each part. The matching error is controlled to be less than 0.01 miles.

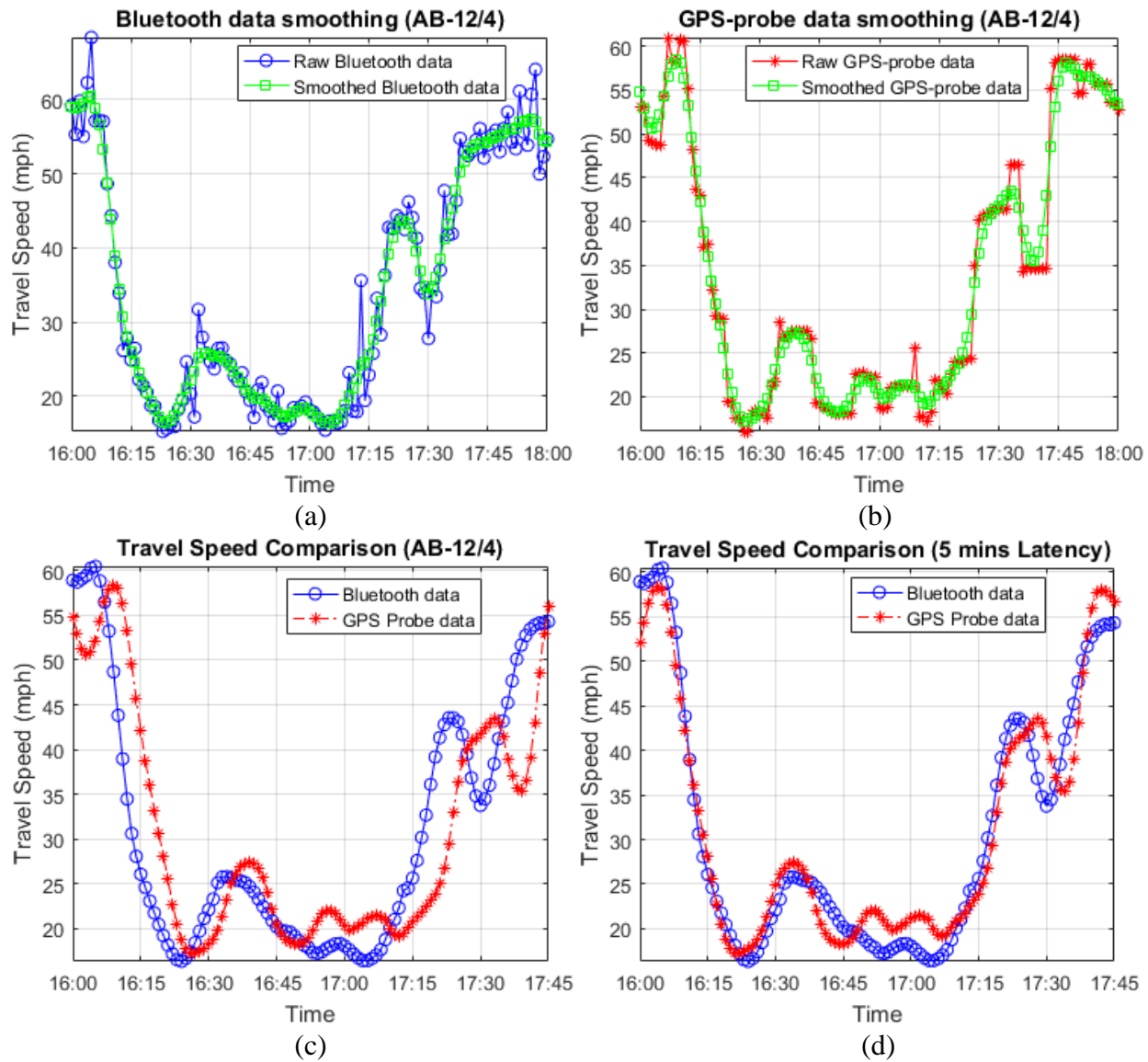

**FIGURE 4 Example of travel speed smoothing and latency measurement (5 minutes) of segment AB during afternoon peak at Dec 4th 2015 (a: Bluetooth data smoothing; b: probe data smoothing; c: original Bluetooth and probe data comparison; d: travel speed comparison of Bluetooth data and latency compensated probe data).**

TABLE 1 shows the average latency calculated based on all identified slowdown episodes, for both morning peak periods and afternoon periods, and based on all three different fitness objectivizes. It can be seen that the average latency measured by three different fitness objectives are really similar to each other, which demonstrates the effectiveness of this methodology and that average latency measurement is "converged". Therefore, this paper uses the average of the latency calculated by three objectives as the GPS-probe data latency. The average probe data latency is 4.26 minutes in the morning peak periods and 3.94 minutes in the afternoon peak periods. Morning and afternoon peak cases combined, for these segments at identified episodes, probe data has an average latency of 4 minutes. Although latency at morning peak is slightly higher than that in the afternoon, there is no significant difference.

**TABLE 1 Average latency and evaluation**

| Period | Number of Observations | Average Latency (minute) | | | |
|---|---|---|---|---|---|
| | | *f1 (AVD)* | *f2 (SVD)* | *f3 (COR)* | *Average* |
| Morning | 32 | 3.96 | 4.42 | 4.41 | 4.26 |
| Afternoon | 45 | 3.64 | 4.01 | 4.19 | 3.94 |

Several graphical example of afternoon latency comparison of original probe data and shifted probe data are shown in FIGURE 5. In the first row, the blue curve represents the Bluetooth data, with the segment name and date shown on top and the original probe data shown as the green curve. Graphs on the second row, show the Bluetooth data and shifted probe data, with resulting latency offset shown on top. The horizontal axis on both graphs shows the time of day, and the vertical axis is travel speed in MPH. Additional rows in this figure show more examples for other segments of different days.

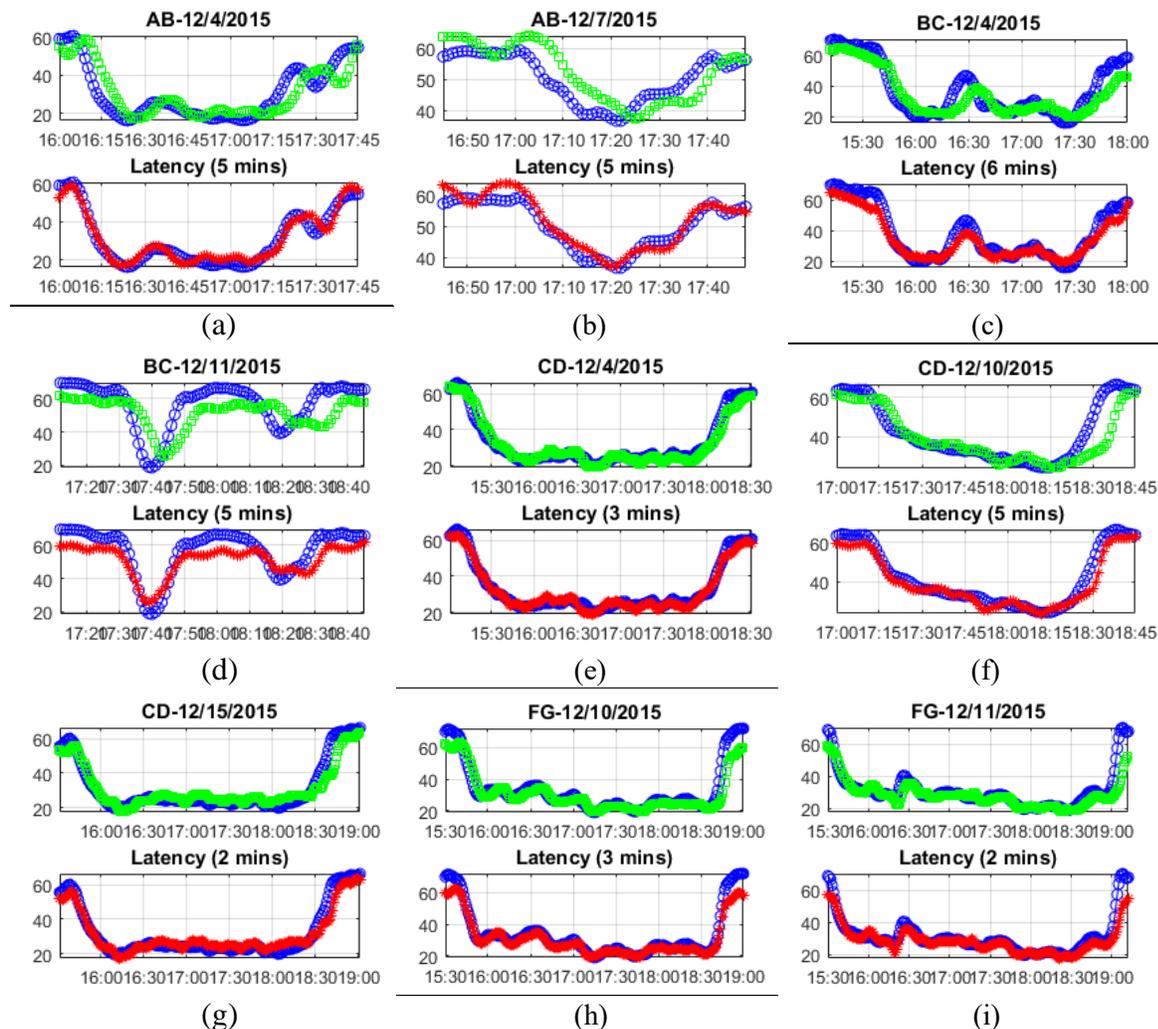

**FIGURE 5 Comparison of original GPS-probe data and shifted GPS-probe data against Bluetooth data**

FIGURE 6 shows the probe data latency distribution. Ass illustrated, the 4 minutes has the highest latency distribution density and the latency distribution is roughly symmetric. The morning peak figure (FIGURE 6a) and afternoon figure (FIGURE 6b) have similar distributions, which further proofs the similarity of latency at morning peak periods and afternoon peak periods. FIGURE 6(c) is the cumulative latency distribution, which demonstrates that 95% of latency values fall within 6 minutes for both morning peak periods and afternoon peak periods.

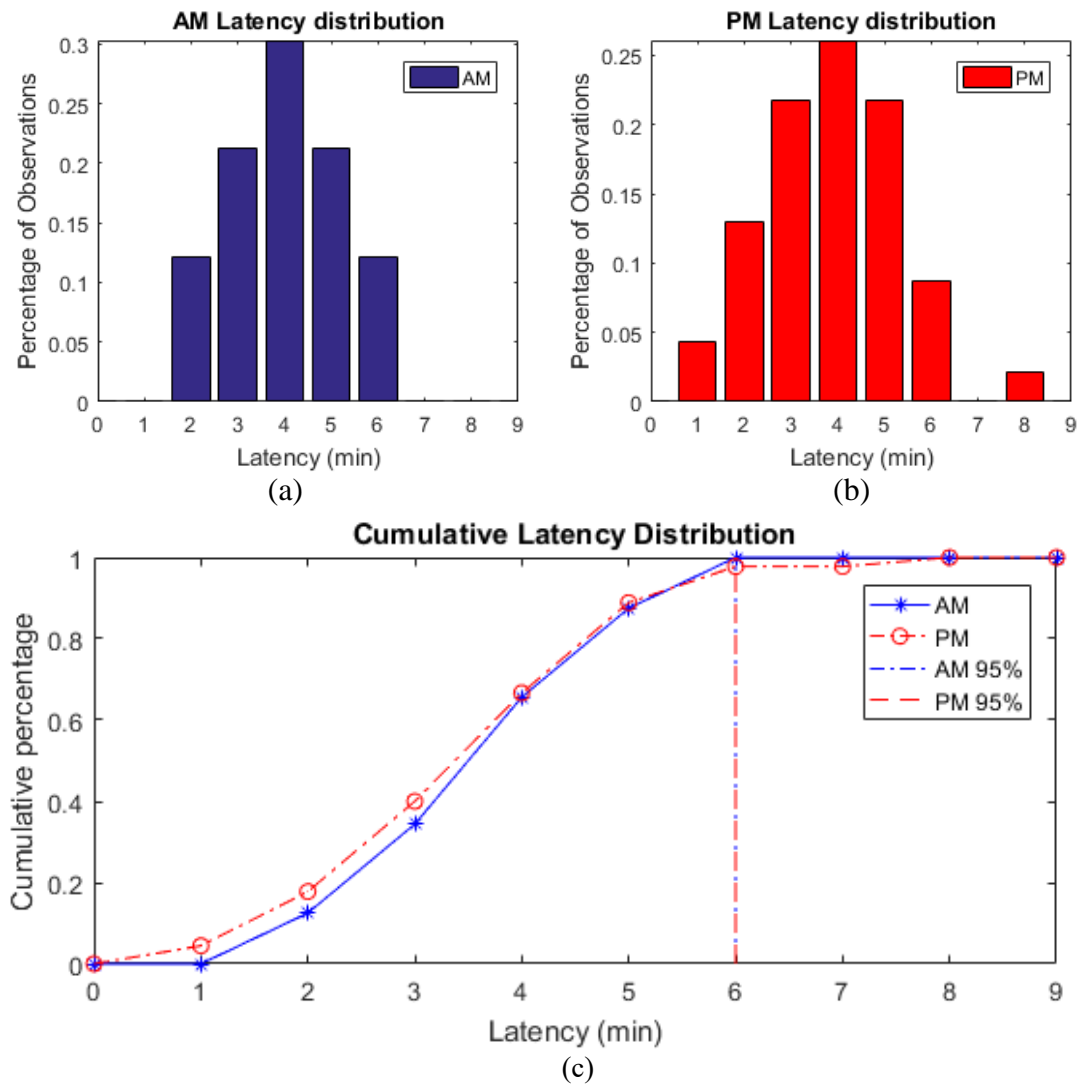

**FIGURE 6 GPS-probe data latency distribution (a: latency distribution at morning peak periods; b: latency distribution at afternoon peak periods; c: cumulative distribution of GPS-probe data)**

**Latency on Different Segments**
Latencies are also calculated on different segments to investigate the impact of segment length on latency. TABLE 2 shows the GPS-probe latencies on different segments ranging from 1.17 miles to 2.20 miles. The average latency is 4.31 minutes, which is consistent with the probe data latency from the previous analysis. The latencies for different segments vary in a small range. The length of TMC segment does not seem to have a significant impact on probe data latency in this study. The scattered plot of latency points in FIGURE 7 also shows that the latency is not significantly correlated with the length of the segment. However, in rural areas some TMC segments might be a lot longer than what it is used in this research which may influence the latency.

**TABLE 2 Probe data latency at different segments**

| Segment | Length (mile) | Average Latency (minute) | | | |
| --- | --- | --- | --- | --- | --- |
| | | f1 (AVD) | f2 (SVD) | f3 (COR) | Average |
| BC | 1.17 | 4.80 | 5.00 | 5.00 | 4.93 |
| KL | 1.28 | 4.43 | 4.86 | 5.00 | 4.76 |
| LM | 1.60 | 3.33 | 3.83 | 3.83 | 3.66 |
| OP | 1.64 | 4.67 | 5.00 | 5.00 | 4.89 |
| AB | 1.69 | 4.56 | 4.56 | 4.67 | 4.60 |
| PQ | 1.70 | 4.78 | 4.89 | 4.89 | 4.85 |
| MN | 1.78 | 4.00 | 4.18 | 3.95 | 4.04 |
| GH | 2.02 | 3.40 | 3.40 | 3.00 | 3.27 |
| CD | 2.07 | 3.92 | 4.50 | 4.50 | 4.31 |
| FG | 2.20 | 2.76 | 4.06 | 4.59 | 3.80 |
| **Average for all segments** | **1.72** | **4.06** | **4.43** | **4.44** | **4.31** |

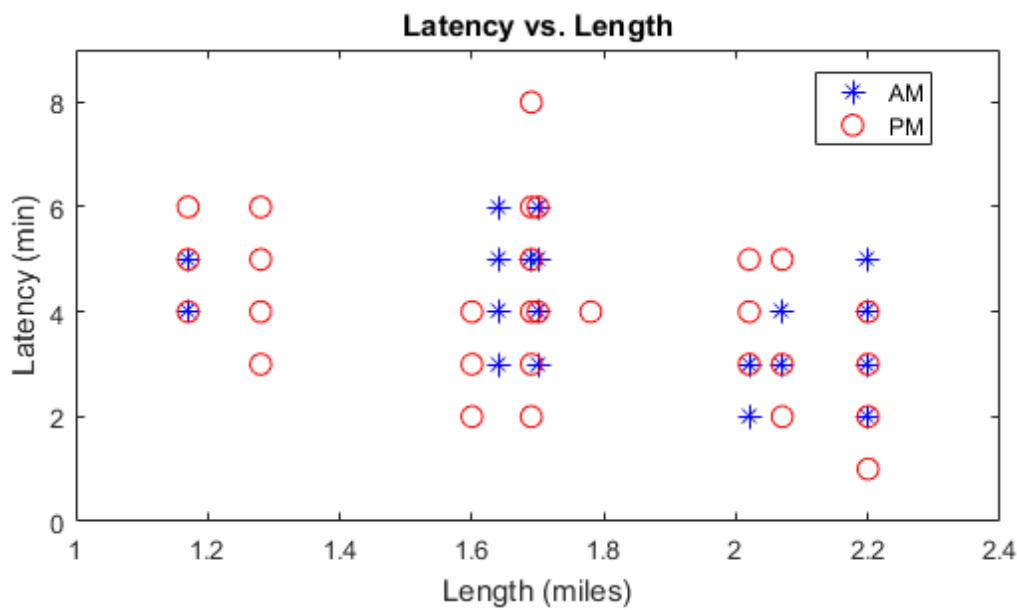

**FIGURE 7 Scattered plot of probe data latency at different length segments**

**Latency at Slowdown and Recovery**
Observations from the empirical work suggest that the latency of GSP-probe data seems to be asymmetrical when comparing the speed reduction period of a slowdown episode to the speed recovery period of the same episode. An example is provided in FIGURE 8 for segment CD during afternoon peak period. As illustrated, the slowdown episode in broken into two parts. The first part starts when the speeds start to decline, and ends at the transition time when speeds start to recover. The transition time is the time that speeds have reached a minimum value during the episode. It must be noted that since the Bluetooth data is used as reference, the minimum speed used to determine the transition time corresponds to the Bluetooth data. The second part of the slowdown episode starts at the speed transition point, and ends when speed fully recovers to the free flow level. To test the asymmetrical latency hypothesis, a total of 77 speed slowdown episodes were analyzed. For each slowdown episode, the start time of the slowdown, the time of the minimum speed, and the end time of the episode when speeds recover were identified. The same methodology discussed previously in the paper was applied to the slowdown and recovery parts of the data separately. TABLE 3 shows a summary of the results for all 77 cases. It can be observed that in general, probe data exhibits smaller latency in capturing reduction of speeds both in morning and afternoon peak. The average latency for capturing slowdown is 3.68 minutes compared to 4.83 minutes for capturing the speed recovery during the morning peak. Similarly in the afternoon peak, probe data captures slowdown with 3.54 minute latency which is lower than 4.76 minute of latency for capturing the speed recovery. In general, the probe data has shown 3.60 minutes of latency for capturing the speed slowdown compared to 4.79 minutes for capturing the speed recovery when all 77 cases are considered as shown in the last row of TABLE 3. In other words, significant reduction in traffic speed seems to be reflected in probe data with 25% less latency compared to the recovery from slowdowns.

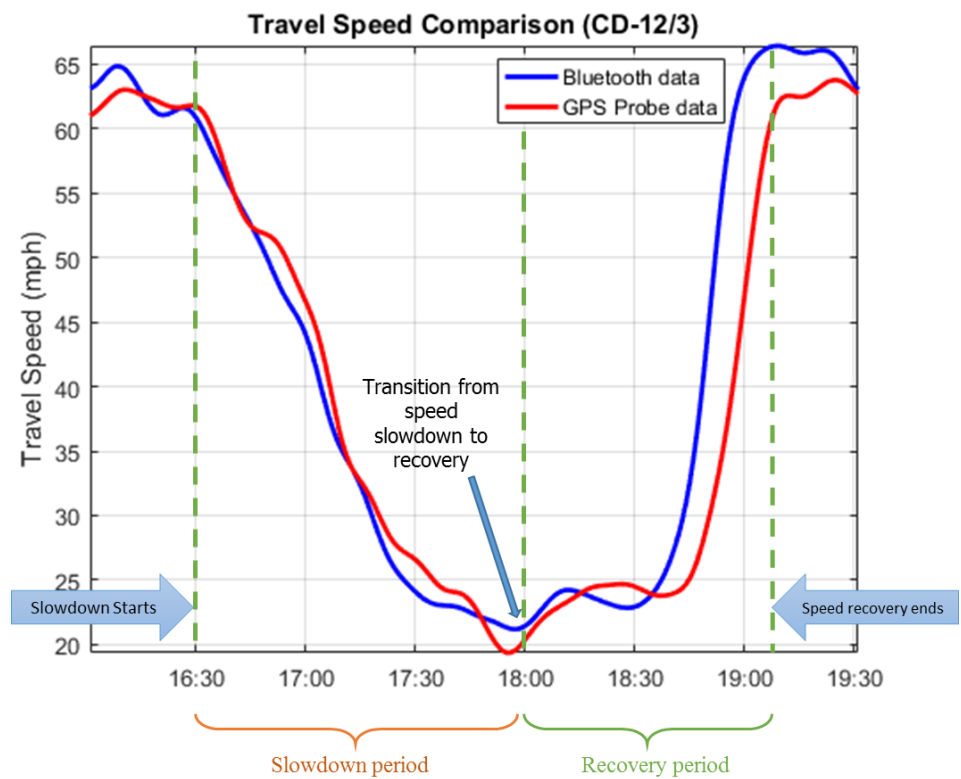

**FIGURE 8 Example of asymmetry of probe data latency during travel speed slowdown and recovery periods**

Wang, Hamedi, Young                                                                                15**TABLE 3 Latency at slowdown and recovery**

| Time Period | Scenario | Number of Observations | Average Latency (minute) | | | |
|---|---|---|---|---|---|---|
| | | | f1 (AVD) | f2 (SVD) | f3 (COR) | Average |
| Morning | Slowdown | 32 | 3.55 | 3.60 | 3.90 | 3.68 |
| | Recovery | 32 | 4.76 | 5.15 | 4.45 | 4.83 |
| Afternoon | Slowdown | 45 | 3.43 | 3.45 | 3.75 | 3.54 |
| | Recovery | 45 | 4.70 | 4.94 | 4.62 | 4.76 |
| Overall | Slowdown | **77** | **3.48** | **3.51** | **3.81** | **3.60** |
| | Recovery | **77** | **4.72** | **5.03** | **4.55** | **4.79** |

# CONCLUSION

This paper makes an effort to analyze and quantify latency associated with GPS probe data compared to Bluetooth traffic data. Several data cleaning and processing methods are described to prepare data step by step. After interpolating data and smoothing the time series, an iterative procedure is discussed to calculate the latency by finding the time shift that maximizes the overlapping of Bluetooth and GPS probe data based on three different fitness objectives. Two freeway corridors were selected to conduct the case study and test the methodology. Results of case study show that the methodology has been successful for measuring the latency of the probe data. It is shown that the latency of probe data in capturing slowdowns is less compared to the latency for capturing speed recovery. The length of the segment does not seem to impact the latency value in the studies scenarios.

Further research is required to investigate the impact of smoothing method on latency measurement. The methodology is robust only if short term volatility in traffic pattern is limited, and thus cannot be applied to measure latency on arterials with high speed fluctuations due to signal timing and mid-block friction factors. Further research is required to design and apply pattern matching algorithms to such cases. The beginning, transition and end time of each slowdown episode in this study were identified manually using a combination of visual graph inspection and statistical analysis which is very tedious. Authors are working towards a methodology for automatic identification of showdowns and their characteristics that is crucial for applying the latency assessment approach on future case studies. Obtaining data from multiple probe data vendors and analyzing latency on different freeways facilities is also subject of future research.

*The results presented in this paper are based on data that is limited in time and scope, and thus they are not conclusive and do not represent the state of latency in commercial probe data in general.* Nonetheless the methodology is proven to be effective and a step in right direction for quantifying the latency of the GPS-probe data.

# ACKNOWLEDGEMENTS
Data used in this study was collected by the I-95 Corridor Coalition as part of their Vehicle Probe Project. The results and conclusions in this document are those of the authors and not the I-95 Corridor Coalition.# REFERENCES
1.   Haghani, A, et al. "Data collection of freeway travel time ground truth with Bluetooth sensors." Transportation Research Record: Journal of the Transportation Research